\documentclass[iop]{emulateapj}  

\shorttitle{Turbulent spectrum at kinetic scales}
\shortauthors{Alexandrova et al.}

\begin{document}




\title{Solar wind turbulent spectrum at plasma kinetic scales}
\author{O. Alexandrova\altaffilmark{1},  C. Lacombe\altaffilmark{1}, A. Mangeney\altaffilmark{1}, R. Grappin\altaffilmark{2,3} and M. Maksimovic\altaffilmark{1}}

\altaffiltext{1}{LESIA-Observatoire de Paris, CNRS, UPMC Universit\'e Paris 06, Universit\'e Paris-Diderot, 5 place J.~Janssen, 92190 Meudon, France.}
\altaffiltext{2}{LUTH-Observatoire de Paris, CNRS, Universit\'e Paris-Diderot, 5 place J.~Janssen, 92190 Meudon, France}
\altaffiltext{3}{Ecole Polytechnique 91128 Palaiseau, France.}

\date{November 15, 2012}

\begin{abstract}

The  description  of the turbulent spectrum of magnetic fluctuations in the solar wind in the kinetic range of scales is not yet completely established. Here, we perform a statistical study of 100 spectra measured by the STAFF instrument on  the Cluster mission, which allows to resolve turbulent fluctuations from ion scales  down to a fraction of electron scales, i.e.  from $\sim 10^2$~km to $\sim 300$~m. We show that  for $k_{\perp}\rho_e \in[0.03,3]$ (that corresponds approximately to the frequency in the spacecraft frame $f\in [3,300]$~Hz), all the observed spectra can be described by a general law $E(k_\perp)\propto k_\perp^{-8/3}\exp{(-k_\perp \rho_e)}$, where $k_{\perp}$ is the wave-vector component normal to the background magnetic field and $\rho_e$ the electron Larmor radius.  This exponential tail found in the solar wind seems compatible with the Landau damping of magnetic fluctuations onto electrons.

\end{abstract}

\section{Introduction}

In neutral, homogeneous and isotropic fluids, the turbulent fluctuations are unpredictable, but their statistics are predictable and universal \citep{frisch95}; the  turbulent spectra follow the power-law $\sim k^{-5/3}$ for any local conditions ($k$ being the wave number). This empirical result was explained by  Kolmogorov \citep{Kolmogorov1941} assuming self similarity of the turbulent  fluctuations between the energy injection scale and the dissipation one $\ell_d$.

In the magnetized solar wind, collisions are very rare (the mean free path is of the order of 1~AU); the dissipation process at work and the dissipation length are not known  precisely. Moreover, in a magnetized plasma, it is difficult to imagine self-similarity over all scales where turbulent fluctuations are observed, since there exist several spatial and temporal characteristic scales, such as the ion Larmor radius $\rho_{i}=\sqrt{2kT_{i\perp}/m_i}/(2\pi f_{ci})$,  the ion inertia length $\lambda_{i}=c/\omega_{pi}$,  the corresponding electron scales $\rho_{e}, \lambda_{e}$, and the ion and electron cyclotron frequencies $f_{ci}$, $f_{ce}$. At these scales,  the dominant physical processes change,  which affects the scaling of the energy transfer time and furthermore the energy transfer rate,  leading to spectral shape changes. 

The first clear spectral change appears at  ion scales.  At 1~AU, the ion scales are nearly equal, $\lambda_i\simeq \rho_i \simeq V/2\pi f_{ci}$,   so it is difficult to determine which of these scales is responsible for the ion break. Independent measurements at different distances from the Sun,  between 0.3 AU to 0.9 AU \citep{bourouaine12}, and a statistical study at 1~AU  \citep{Leamon2000}  indicate that the spectral break is related to the ion inertia length $\lambda_i$.  Nearly incompressible magnetic fluctuations cascading from the inertial range may undergo kinetic  effects in the vicinity of the ion scales. At these scales, ion temperature anisotropy instabilities occur \citep{gary01}  and can remove or inject energy in the turbulent cascade.    However, for most of the solar wind observations, the plasma is stable  \citep{matteini07,matteini11,bale09}. The energy re-distribution among the fluid and kinetic degrees of freedom  in the vicinity of ion scales is still a matter of debate and is probably at the origin of the spectral variations observed between 0.3  and 3~Hz in the satellite frame: the spectral index here varies between $-4$ and $-2$  \citep{Leamon1998,smith06,sahraoui10}.  This spectral range is usually attributed to the ion dissipation range \citep{Leamon1998,Leamon1999,Leamon2000,smith12}  or to another fluid cascade, which may continue down to electron scales \citep{Biskamp96,Stawicki2001,Li2001,Galtier2003b,GaltierBuchlin2007ApJ}.

Between ion and electron scales,  the fluctuations of the electron fluid form a small scale inertial range \citep{alexandr07,alexandr08apj}, or, following the nomenclature of \citet{smith12}, an electron inertial range. Here, indeed, a reproducible spectrum $\propto k_{\perp}^{-2.8}$ is observed \citep{alexandr09,chen10,sahraoui10}. Approaching electron scales, one may expect to observe an electron dissipation range, as was suggested by  \citet{alexandr08npg}. At such small scales,  there are only a few observations \citep{alexandr09,sahraoui10} and  the descriptions are different. Larger statistical studies are needed to establish more firmly the properties of turbulent spectra at electron scales.

In this paper  we present a large statistical study of magnetic spectra starting at ion scales and going beyond electron spatial scales. We use data from the STAFF instrument~\citep{cornilleau97} on the Cluster mission~\citep{Escoubet1997}, which is able to measure such a range of scales.  In a previous study, \citet{alexandr09} described   the electron inertial and the electron dissipation ranges separately:   a power-law $\sim k^{-2.8}$ for the inertial range  and a curved spectrum $\propto\exp(-\sqrt{k/k_0})$ for the dissipation range. This model is rather complicated and has a large number of free parameters. In the present study, we propose  a single algebraic description for both ranges,  namely, an exponential with a power-law pre-factor: $E(k_{\perp})=Ak_{\perp}^{-\alpha}\exp(-k_{\perp}\ell_d)$. We find that this model describes well the totality of the observed spectra  at scales smaller than $\lambda_i$ and $\rho_i$  and that its cut-off scale $\ell_d$ correlates with~$\rho_e$. The power-law exponent $\alpha$ is found to be close to $-8/3$. This model  (henceforth called ``the  {\it exp}--model")  has only one free parameter, the amplitude of the spectrum.

Previous authors \citep{sahraoui10} have used a double power-law model with a break to fit the observations  in the electron inertial and dissipation ranges. We  have applied  this model as well to our data,  and we find that the first power-law exponent is consistent with the previous studies \citep{alexandr09,chen10} while the second exponent varies a lot. Despite the fact that the double power-law model has  more free parameters than the exponential model used here,  we find that it describes only 30\% of the observed spectra and that the associated break scale does not present any clear correlation with an electron characteristic scale.

\section{Observations}

\begin{figure}
\includegraphics[width=8.0cm]{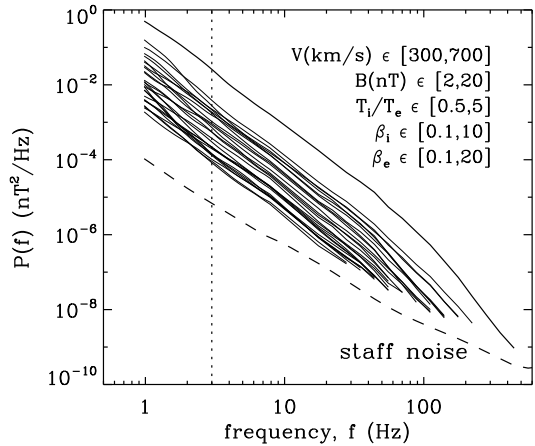}
\caption{Frequency spectra with a signal to noise ratio  greater than 3 measured by Cluster--1/STAFF in the free solar wind (for 27 intervals randomly chosen among 100). The dashed line shows the instrument noise level. The vertical dotted line corresponds to $f=3$~Hz.  The legend indicates the variations of some solar wind parameters for the studied data set of 100 spectra:  speed $V$, magnetic field $B$,  temperature ratio and the ion and electron plasma $\beta$. }
\label{fig:spec102}
\end{figure}
\begin{figure}
\includegraphics[width=9.0cm]{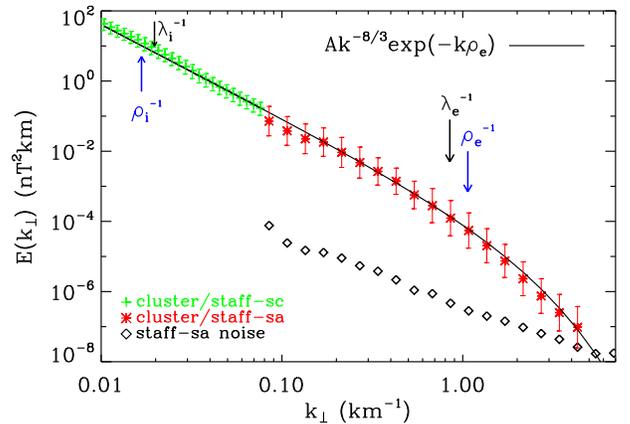}
\caption{ Fit of the most intense spectrum of Figure~1 with the {\it exp}--model. The spectrum was measured by Cluster-1/STAFF on 22/01/2004.   Green crosses represent the SC measurements, red stars show the raw SA measurements without correction of the 1st three underestimated points, visible here around $0.1$km$^{-1}$. Diamonds indicate the STAFF-SA noise level. The blue arrows indicate inverse ion and electron Larmor radii, the black ones  correspond to the inertial lengths. The solid line gives the exp-model $Ak^{-8/3}\exp(- k\rho_e)$.  }
\label{fig:spec66}
\end{figure}

For our statistical study we select homogeneous intervals of 10 minutes (long enough to study kinetic scales)  within the five years interval (2001--2005) of Cluster. We eliminate  time intervals during which Cluster is magnetically connected to the bow-shock by using electrostatic wave spectrograms, which show clearly waves typical of the electron foreshock  \citep{etcheto84,lacombe85}, and by using the shock model described by \citet{Filbert1979}.   For small angles $\Theta_{BV} $ between the interplanetary magnetic field ${\bf B}$ and the solar wind velocity ${\bf V}$, Cluster is connected to the shock. Thus, our data set only contains intervals for which the angle $\Theta_{BV} > 60^{\circ}$. If the turbulent  fluctuations have a phase speed $V_{\phi}\ll V$,  Cluster detects  by Doppler shift the fluctuations with ${\bf k\| V}$. As ${\bf B}$ and ${\bf V}$ are quasi-perpendicular,  Cluster measures fluctuations with ${\bf k\perp B}$. We apply the Taylor hypothesis to get the wave-number from the frequency, $k_{\perp}=2\pi f/V$. However, about $\sim 10\%$ of the pre-selected intervals show the presence of   right hand polarised whistlers in quasi-parallel propagation. For these waves  the Taylor hypothesis is not applicable, because $V_{\phi} > V$. We  discard these intervals in the present study. This data selection process gives us 100~intervals.  Within this statistical sample, the plasma conditions vary as usually in the solar wind in fast and slow streams at 1~AU (see the legend of Figure~\ref{fig:spec102}).

Figure~\ref{fig:spec102} shows the total Power Spectral Density (PSD) of magnetic fluctuations, for  27 intervals randomly chosen among 100, as a function of frequency in the spacecraft frame $P(f)$. 
These spectra are measured by the STAFF Search Coil sensor and analyzed onboard by the magnetic waveform unit (hereafter called SC) at $f\in[0.5,9]$~Hz  and by the Spectrum Analyser (hereafter called SA) at $f\geq 8$~Hz.  

The spectra are analyzed only for the frequencies where the Signal to Noise Ratio (SNR) is larger than~3. The spectral parts below this threshold are not shown to avoid any  erroneous interpretation. As one can see from Figure~\ref{fig:spec102}, this instrumental noise limit allows us to use data up to $30-400$~Hz, depending on the turbulence intensity (i.e., for the most intense spectrum, we have valid observations up to $400$~Hz).  The analyzed range of frequencies corresponds to $f\in ]f_{ci},f_{ce}]$.
 
A poor calibration of the first 3 frequencies of SA (at 8, 11 and 14~Hz) [Y. de Conchy and N. Cornilleau, private communication, 2011], was corrected by an interpolation of these points between the highest SC frequency and the 4th point of the SA spectra. The linear interpolation between $\log_{10} P(f)$ and $\log_{10} f$ is possible as far as the spectra follow a power-law at these frequencies.  An example of a raw spectrum without the correction can be found in Figure~\ref{fig:spec66}.

\section{Algebraic description of turbulent spectra at scales smaller than $\rho_i$ and $\lambda_i$.}
\subsection{Exponential model}

Here we propose a model to describe the whole turbulent spectrum  at scales smaller than $\rho_i$ and $\lambda_i$ and down to a fraction of the electron scales with the  smaller  possible number of  parameters, namely an exponential  with a characteristic scale $\ell_d$ and with a power-law pre-factor
\begin{equation}\label{eq:exp}
E(k_{\perp})=A k_{\perp}^{-\alpha}\exp(- k_{\perp}\ell_d).
\end{equation}
This {\it exp}--model has three free parameters: the amplitude $A$, the spectral index $\alpha$ and the cut-off or "dissipation" scale~$\ell_d$.

We start by fitting the model (\ref{eq:exp}) to the 100 observed spectra (with a signal to noise ratio $>3$, as explained in section~2) for $k_{\perp}$ corresponding to $f>3$~Hz (see vertical dotted line in Figure~\ref{fig:spec102}), assuming that the three parameters have independent variations.  

Figure~\ref{fig:spec66} gives  the fit with the most intense spectrum of Figure~\ref{fig:spec102} as a function of the wave-number $P(k_{\perp})=P(f)V/2\pi$, which is determined using the Taylor hypothesis and the energy conservation law $\int P(k_{\perp})dk_{\perp} = \int P(f)df$. Green crosses show the Morlet wavelet spectrum \citep{Torrence1998} of STAFF-SC measurements. Red stars display the STAFF-SA data for the same time period. (In this plot we keep the 3 first poorly calibrated data points, one can see them around $k=0.1$~km$^{-1}$ and compare with the result of the interpolation in Figure~1).  The error bars are estimated from the variance  over 10 minutes of the PSD at each frequency \citep{alexandr10}. This spectrum is valid up to $\simeq 400 Hz$, which gives us the maximum wave-vector $k\sim 4$~km$^{-1}$ (while $1/\rho_e \simeq 1$~km$^{-1}$).  This is the smallest scale ever measured with a good sensitivity at 1~AU in  the solar wind. The exp-model (\ref{eq:exp}) fitting is shown by the black solid line.  The parameters of the fit in this case are $\alpha = 2.70\pm 0.15$ and $\ell_d = (0.90 \pm 0.25)$~km, while $\rho_e=(0.95\pm 0.05)$~km.

\begin{figure}[htb] 
\includegraphics[width=8.4cm]{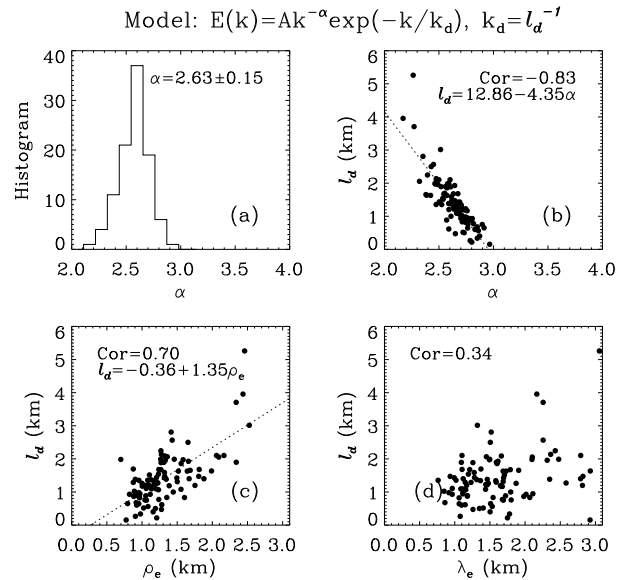}
\caption{ Results of the fitting with the {\it exp}--model for the 100 observed spectra: (a) histograms on the spectral index $\alpha$; (b) cut-off scale $\ell_d$ as a function of $\alpha$; (c) $\ell_d$ as a function of the electron Larmor radius $\rho_e$ and (d) $\ell_d$ as a function of the electron inertial length $\lambda_e$.}
\label{fig:fitexp}
\end{figure} 

Figure~\ref{fig:fitexp} summarizes the results of the fitting for the 100 spectra. Panel (a) shows the histogram of the spectral index, $\alpha=2.63 \pm 0.15$,   the error being the standard deviation of the mean. Note that $\langle\alpha\rangle \simeq 8/3$.  
 It appears that the variations of $\alpha$ and $\ell_d$ are not independent since  the dispersion in $\alpha$ is due to the  variations of the cut-off scale $\ell_d$ as observed in Figure~\ref{fig:fitexp}b.  A linear fit gives $\ell_d {\text (km)}=12.9-4.4\alpha$, i.e.
\begin{equation}\label{alpha-Ld}
\alpha=2.9-\ell_d/4.4 
\end{equation}
i.e.  if $\ell_d$  was small, $\alpha$ would be approximatively equal to $2.9$, a value close to the one found by \citet{alexandr09}.

On the other hand, the variations of $\ell_d$ are related to the variations of  the electron Larmor radius, $\ell_d\sim 1.35 \rho_e$, as shown in  Figure~\ref{fig:fitexp}c, with a relatively high  correlation coefficient of $0.70$. Figure~\ref{fig:fitexp}d shows a positive but much weaker correlation of $0.34$ between the dissipation scale $\ell_d$ and the electron inertia length   $\lambda_e$. 

The  results presented in Figure~\ref{fig:fitexp} suggest that within  the framework of  the exponential model there is  only one free parameter, the amplitude of the turbulent spectra, $A$, and the observed spectra can be described approximately by  
\begin{equation}
E(k_{\perp}) \simeq Ak_{\perp}^{-8/3}\exp(-k_{\perp}\rho_e). 
\end{equation}
\begin{figure}[htb] 
\includegraphics[width=8.6cm]{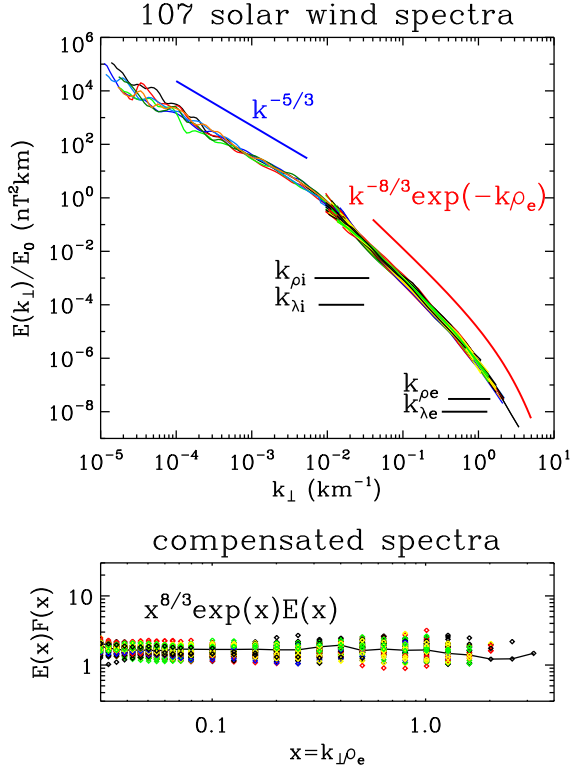}
\caption{Upper panel: 107 superposed solar wind spectra, the 100 spectra in the kinetic range analysed in this paper and the 7 spectra covering fluid and kinetic scales from  \citep{alexandr09}, the constant being $E_0\in[0.4,95]$. All spectra have the signal to noise ratio greater than 3, as in Figure~\ref{fig:spec102}.  The blue line indicates $\propto k_{\perp}^{-5/3}$ and the red line shows $\propto k_{\perp}^{-8/3}\exp(-k_{\perp}\rho_e)$, that represents well all the spectra. Bottom panel:  the diamonds show 100 compensated spectra $E(k_{\perp}\rho_e)\cdot F(k_{\perp}\rho_e)$, with $F(k_{\perp}\rho_e)=(k_{\perp}\rho_e)^{8/3}\exp(k_{\perp}\rho_e)$, for $k_\bot \rho_e \geq 0.03$. The solid line denotes the  most intense compensated  spectrum.}
\label{fig:5}
\end{figure} 
We verify this point in Figure~\ref{fig:5}, where we superpose  the 100 spectra analyzed here with the 7 spectra  covering fluid and kinetic scales from  \citep{alexandr09}.  The spectra are shifted vertically by a parameter $E_0$ (a relative spectral level), in the same way as in Fig.~2 of \citep{alexandr09}. The superposition of the 107 spectra is nice, which indicates the    generality of the turbulent spectrum $E(k_{\perp})$ in the solar wind: it follows $\propto k_{\perp}^{-5/3}$ at MHD scales and $\propto k_{\perp}^{-8/3}\exp(-k_{\perp}\rho_e)$ at scales smaller than the ion kinetic scales $\lambda_i$ and $\rho_i$ (i.e. $k\rho_e \geq 0.03$).  The bottom panel  shows the 100  spectra $E(k_{\perp}\rho_e)$ compensated by a function $F= (k_{\perp}\rho_e)^{8/3}\exp(k_{\perp}\rho_e)$  for $k\rho_e\geq 0.03$:  the resulting spectra are flat, indicating that the  {\it exp}--model with one free parameter describes well all the turbulent spectra in the solar wind at these scales and is valid for nearly 2 decades in wave numbers.  Note that a damping length $\ell_d$ variation of more than  20\% with respect to the mean $\rho_e$ values results in a strong departure of the compensated spectra.   
  
The amplitude $A$  of the turbulent spectra (related to the  parameter $E_0$) is  found to be correlated with the ion thermal pressures $nkT_i$, as within the MHD range of turbulence \citep{Grappin90}, and with the ion temperature anisotropy $T_{i\perp}/T_{i\|}$ (not shown).  Other plasma parameters seem to be less important, but still it is impossible to exclude completely the influence  of the magnetic and kinetic energies in the solar wind (paper in preparation).

\subsection{Break model}

Is there another simple model which represents well the observations with a small number of free parameters?  Let us compare the turbulent spectra  within the electron inertial and dissipation ranges (i.e. scales smaller than $\rho_i$ and $\lambda_i$) with  the  double power-law or {\it break}-model
\begin{equation}\label{eq:break}
\tilde{E}(k_{\perp})=A_1k_{\perp}^{-\alpha_1}(1-H(k_{\perp}-k_b)) + A_2k_{\perp}^{-\alpha_2}H(k_{\perp}-k_b),
\end{equation}
$H(k_{\perp}-k_b)$ being the Heaviside function, $k_b$ the wave number of the break scale $\ell_b=1/k_b$, $A_{1,2}$ the amplitudes of the two power-law functions with spectral indices $\alpha_{1,2}$ on both sides of $k_b$. This model has five free parameters.  Note that equation (\ref{eq:break}) is not differentiable for $k=k_b$. Near $k_b$ the  turbulent level  has to be the same on both sides. Thus, $k_b$ can be determined by the four other parameters of the model 
\begin{equation}\label{eq:break-1}
\log_{10}k_b=\frac{\log_{10}(A_1/A_2)}{\alpha_1-\alpha_2} .
\end{equation}
Iterations to find the model parameters with condition (\ref{eq:break-1})  converge to different results, depending on the initial $k_b$. Therefore, minimising the error of the fit over $k_b$ is needed, so that finally the model has still five free parameters.  

We  apply this {\it break}--model to the 100 observed spectra within the same $k_{\perp}$--range as was done for the {\it exp}--model.  Despite the fact that the {\it break}--model has more free parameters than the exponential model, we find that it  can be  applied only to 30 spectra;  for the 70 other spectra there is no solution with condition (\ref{eq:break-1}) verified,  indicating the absence of a clear break point, or, for a part of these spectra, not enough data points to isolate the second power-law.

 \begin{figure}[htb] 
\includegraphics[width=8.4cm]{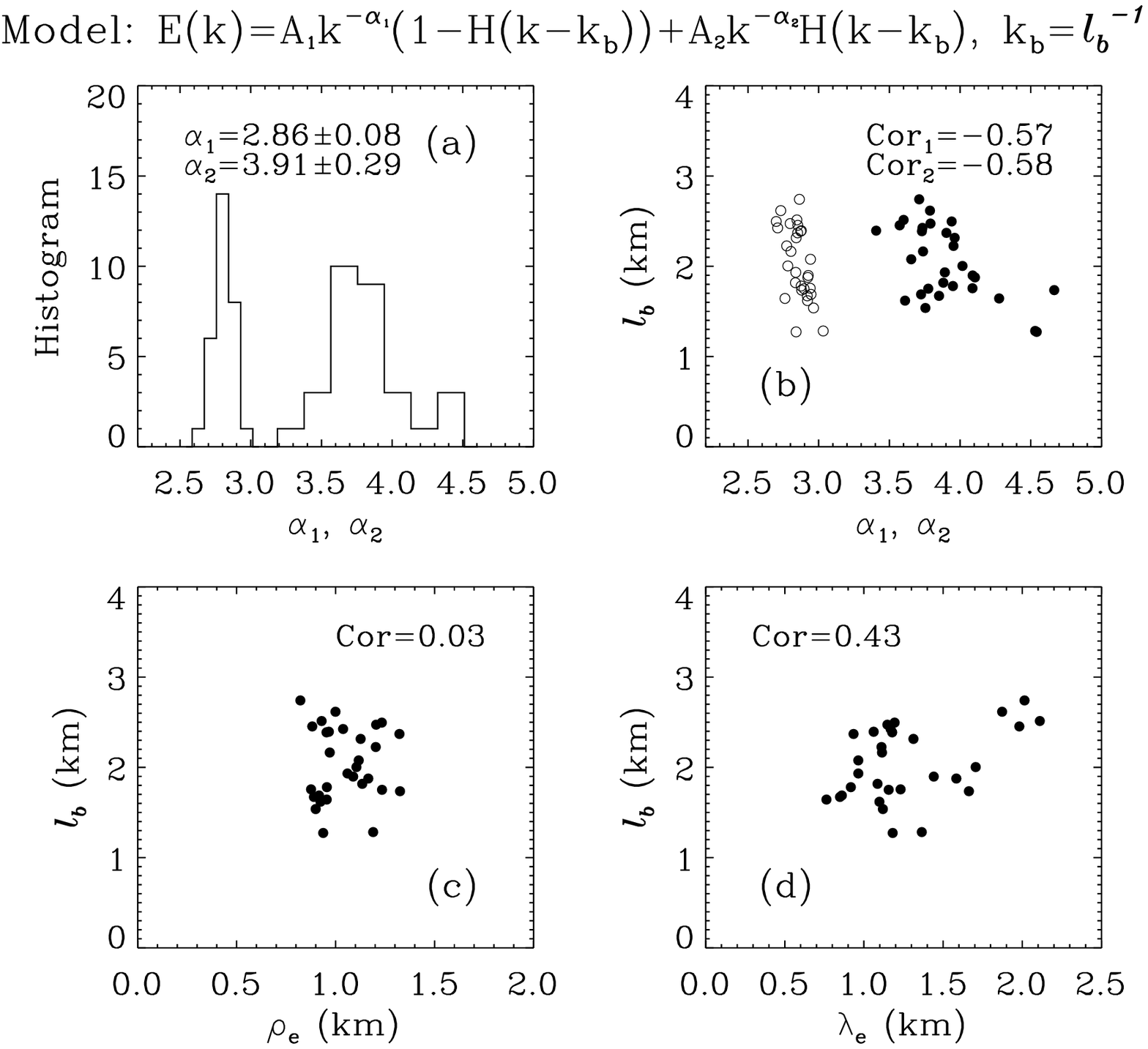}
\caption{\label{fig:break-model-res} same format as Figure~\ref{fig:fitexp}, but for the parameters of the {\it break}--model  determined from the fitting with 30 spectra, i.e. (a) histograms on the spectral indices; (b) break scale $\ell_b$ as a function of $\alpha_1$ (open circles) and $\alpha_2$   (filled circles); (c) $\ell_b$ as a function of the electron Larmor radius $\rho_e$ and (d) $\ell_b$ as a function of the electron inertial length $\lambda_e$.}
\end{figure}
Figure~\ref{fig:break-model-res} summarizes the  parameters of the  {\it break}--model determined from the fitting of the 30 spectra. For these spectra, the condition (\ref{eq:break-1}) is verified. Panel (a) shows histograms of the spectral indices: the mean values are  $\alpha_1 = 2.86\pm 0.08$, $\alpha_2 = 3.91 \pm 0.29$. Note the narrow dispersion of the spectral index $\alpha_1$. It is close to $\alpha$, when $\ell_d$ is negligible (see eq.~(\ref{alpha-Ld}))  so confirming the spectrum between ion and electron scales found by \citet{alexandr09}. So we can fix one of the parameters of the model. The second exponent $\alpha_2$ has a large dispersion, not found to be controlled by any plasma parameter.  The values of the spectral indices are correlated to the position of the break scale $\ell_b$ (see panel b).  Figure~\ref{fig:break-model-res}c shows $\ell_b$ as a function of $\rho_e$. No correlation is observed. Figure~\ref{fig:break-model-res}d shows $\ell_b$ as a function of $\lambda_e$: the correlation is positive but weak ($\simeq0.43$).  From the comparison of the observed turbulence spectra at plasma kinetic scales with the {\it break}--model,  one may conclude that this model has one fixed parameter, another is fixed by the condition (\ref{eq:break-1}), the other 3 parameters are free, not found to be determined by plasma parameters. 

\begin{figure}[htb] 
\includegraphics[width=8.6cm]{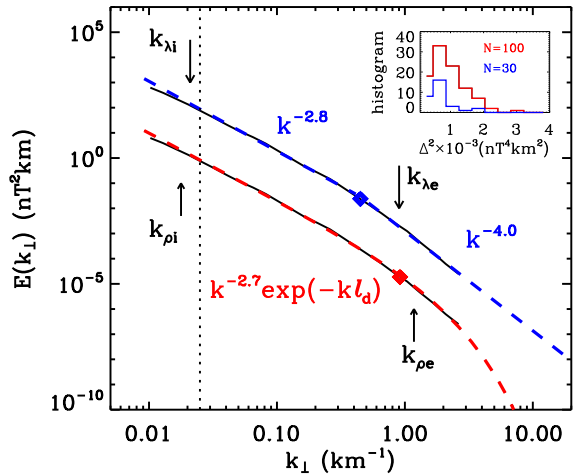}
\caption{Example of an observed spectrum (lower black line)  compared with $E(k)\sim k^{-2.7}\exp(-k\ell_d)$ (red dashed line), with $\ell_d^{-1}$  marked by a red diamond; the error of the fit is $\triangle_{exp}^2=1.8\times 10^{-3}$~(nT$^4$km$^2$).  The same spectrum, shifted by a factor $10^2$ and compared with  $E(k)\sim k^{-2.8}$ below  the break shown by  a blue diamond, and with $\sim k^{-4.0}$ above it (blue dashed line); here $\triangle_{break}^2=1.9\times 10^{-3}$~(nT$^4$km$^2$). Ion and electron characteristic scales are marked by arrows; a vertical dotted line indicates the beginning of the $k$--domain where the fits are performed. Insert: histograms of the fitting errors, $\triangle_{exp}^2$ (red, for 100 spectra) and $\triangle_{break}^2$ (blue, for 30 spectra).
}
\label{fig:event67}
\end{figure}
It is important to note that the errors of the fit of the {\it break} and {\it exp}--models are   of the same order.  Figure~\ref{fig:event67} shows an example of a solar wind spectrum (black lines) fitted with both models,   with the least mean square distance  between the observed spectra and the fit, $\triangle^2$, given in the caption; the insert shows histograms of $\triangle^2_{exp}$  (red) for 100 spectra and $\triangle^2_{break}$ (blue) for 30 spectra: the same mean values of the histograms are observed. This leaves us free to choose the model, based on other criteria than goodness of the fit, namely, on the number of degrees of freedom of the model and on the number of described cases.

\section{Discussion and Conclusion}

The {\it exp}--model $E(k_{\perp}) = A k_{\perp}^{-8/3}\exp(-k_{\perp}\rho_e)$ proposed in this study provides a single algebraic description of the solar wind spectrum at scales smaller than the ion characteristic scales, $\lambda_i$ and $\rho_i$, and going beyond the electron scales (i.e. within the electron inertial and dissipation ranges).  This model describes well the totality of the observed spectra and has only one free parameter -- the amplitude $A$ of the spectrum. The amplitude seems to be a function of the ion thermal pressure and the ion temperature anisotropy in the solar wind. However, it is difficult to exclude the role of the magnetic and kinetic energies: more work is needed to determine the exact relationship between the amplitude of the turbulent spectrum and the energy budget in the solar wind.

 The spectral index close to $-8/3$ observed in the solar wind at scales smaller than ion characteristic scales is in agreement with quasi-bidimentional strong Electron MHD turbulence ($k_{\perp} \gg k_{\|}$) when parallel cascade is weak  \citep{galtier05}. Recently, the same spectral index was found as well in strong kinetic Alfv\'en turbulence  \citep{boldyrev12}.

In usual fluid turbulence, the far dissipation range is described by $E(k) \sim k^{3}\exp(-c k\ell_d)$ (with $c\simeq 7$) \citep{chen93}. 
This is due to the resistive damping rate $\propto k^2$ valid in a collisional fluid, which gives an exponential spectral tail. 
In  the collisionless plasma of the solar wind there is no resistive damping, and thus this  coincidence deserves an explanation. 

\citet{howes11pop} consider a model (``weakened cascade model'') which includes the nonlinear transfer of energy from large to small scales in Fourier space  
and the damping of kinetic Alfv\'en waves (KAW's). 
For completeness, we discuss now this model in some detail. The model reads for the magnetic energy $b^2$ at scale $k$:
\begin{equation}\label{eq:transf}
\partial_t b^2_k = - k_\bot \partial_k \epsilon - 2 \gamma b^2_k + S
\label{mou}
\end{equation}
with $\epsilon$ being the magnetic energy transfer rate, $S$ being the source term.
 The damping term $2 \gamma b^2_k$ is obtained by linearizing the Vlasov-Maxwell equations in the gyrokinetic limit ($k_\| \ll k_\bot$, with frequencies $f \ll f_{ci}$). In the limit $k \rho_i \gg 1$, this gives 
 \begin{equation} \label{eq:gamma}
 \gamma \simeq k_\parallel V_a (k_\bot \rho_i)^2,
 \end{equation}
 see eq.~(63) in \citep{howes06}.
To complete the model, we must write down the expression for the magnetic energy flux $\epsilon$. This is given by 
\begin{equation}
\epsilon = b^2 / \tau = k_\bot u b^2
\label{epsilon}
\end{equation}
where $\tau = 1/k_\bot u$ is the nonlinear time  and $u$ the  velocity fluctuation.  At MHD scales $k_\bot \rho_i \ll 1$, we have for Alfv\'en waves $u \simeq b$, but at small scales $k_\bot \rho_i \ge 1$, we have for KAW $u /b \simeq k_\bot \rho_i$, see eq.~(3) in \citep{howes11pop}. Using these expressions in eq.~(\ref{mou}) and (\ref{epsilon}), one obtains scaling laws for the magnetic energy spectrum as stationary solutions of the transfer equation (\ref{eq:transf}), neglecting the damping and source terms. 
The spectral laws are respectively $E_k \propto k_\bot^{-5/3}$ at large scales and $E_k \propto k_\bot^{-7/3}$ 
between ion and electron scales. When taking into account the damping term, \citet{howes11pop} obtain numerically the same spectral laws, with a final curved tail at scales smaller than electron scales. Superficially, this spectrum thus resembles the analytic form which we have found to be valid to describe  the solar wind turbulence. 

We last remark that the damping term in the model  of \citet{howes11pop}  for $k_\bot \rho_i \gg 1$ is (see eq. (7)) of the form 
$\gamma \propto k_\| k_\bot^2$. 
Taking into account the  assumption of critical balance $\tau=\tau_A$ (i.e. $k_{\perp}u=k_\| V_A$) \citep{Goldreich1995}, with $\tau_A$ the Alfv\'en time and $V_A$  the Alfv\'en speed, and the spectral index $-7/3$ (i.e. $u\sim k_\bot^{-2/3}$),  one gets $k_\| \propto k_\bot^{1/3}$. Therefore, the damping term takes the form  $\gamma \propto  k_\bot^{2+1/3}$.  The exponent of the damping rate is thus very close to the $k^2$ scaling of the Laplacian viscous term, which is known to lead in hydrodynamical turbulence to an exponential tail in the dissipation range.

 This model does not take into account the cyclotron damping. So, while the proposed phenomenology may explain the exponential tail of the $k_{\perp}$--spectrum studied here, it cannot describe more isotropic  wave vectors, which might be present as well in the solar wind.  It is possible that the dissipation mechanism could also be due to electron-cyclotron absorption of oblique short-wavelength whistler waves, or even of lower-hybrid waves. More observations under different field-to-flow angles $\Theta_{BV}$ are needed within the electron inertial and dissipation ranges to address this point. 

To build a realistic model of the dissipation in the solar wind we need to resolve still an open question on the nature of the turbulent fluctuations. Some authors argue that the electron inertial range is a whistler mode turbulence \citep{saito10,narita-gary10}, others suggest KAW turbulence \citep{bale05,Schekochihin2009ApJS,sahraoui10,salem12}, or a combination of both types of linear waves \citep{podesta-gary11}. The model of \citet{howes11pop}, described above is based on KAW turbulence as well.  However, it is still not clear whether we can describe turbulence in the solar wind as a mixture of linear waves (weak turbulence) which will dissipate homogeneously in space (or in the plane perpendicular to ${\bf B}$), or if it is a strong turbulence with dissipation restricted to intermittent coherent structures.  What is the topology of these structures -- current sheets, shocks or coherent vortices? 

In the present study we have limited ourselves to observations of the spectral shape in the electron inertial and dissipation ranges.  Our results give observational constraints for future theoretical models.

\appendix

\section{Plasma parameters variations during the spectra integration time}

\begin{figure}
\includegraphics[width=17.2cm]{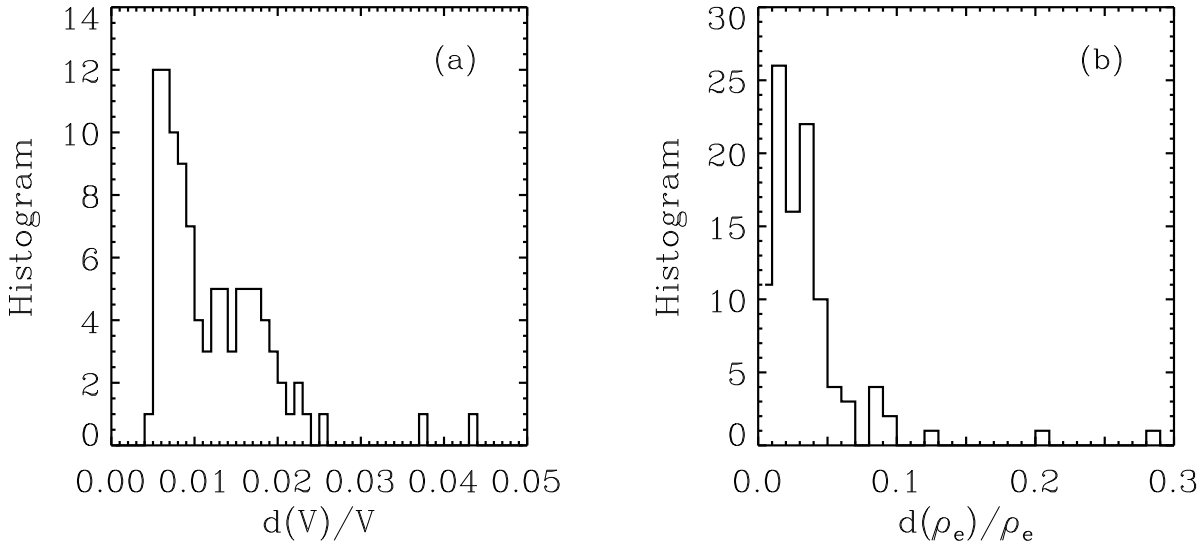}
\caption{\label{fig:a1}   Histogram of the variability of the solar wind speed $V$ during 10 minutes, for a sample of  
100 intervals (left panel). Right panel : histogram of the 
variability of the electron gyroradius $\rho_e$.   } 
\end{figure}

One could argue that the exponential bending found in the $k$-spectra of the 
magnetic fluctuations is due to variations of the solar wind speed $V$ or of  
the gyroradius $\rho_e$ during the 10 minutes of each considered interval. 
Indeed, each $k$-spectrum is obtained with an average $P(f)$ over 150 frequency spectra, itself shifted in the $k$-domain with the average $V$. The standard deviation $d V$ over 10 minutes is very small. Figure~\ref{fig:a1}(a) displays the histogram of the ratio $d V/V$: 91\% of the 100 considered 
intervals have $d V/V < 0.02$. Thus, the shift of $P(f)$ in the k-domain 
with the average $V$ cannot change the spectral bending. Similarly, the  standard deviation $d(\rho_e)$ over 10 minutes is small. Figure~\ref{fig:a1}(b) displays the histogram of the ratio $d(\rho_e)/\rho_e$ : 96\% of the 100 considered intervals have $d(\rho_e)/\rho_e < 0.1$. Thus, the use of  the average $\rho_e$ for an interval, in place of the exact $\rho_e$ for  each of the 150 k-spectra, cannot produce the observed bending which covers a wide range of  $k$, and cannot smooth a possible spectral break. 

\section{Log-spaced frequencies of the Cluster/STAFF-SA instrument}
\begin{figure}
\includegraphics[width=17.2cm]{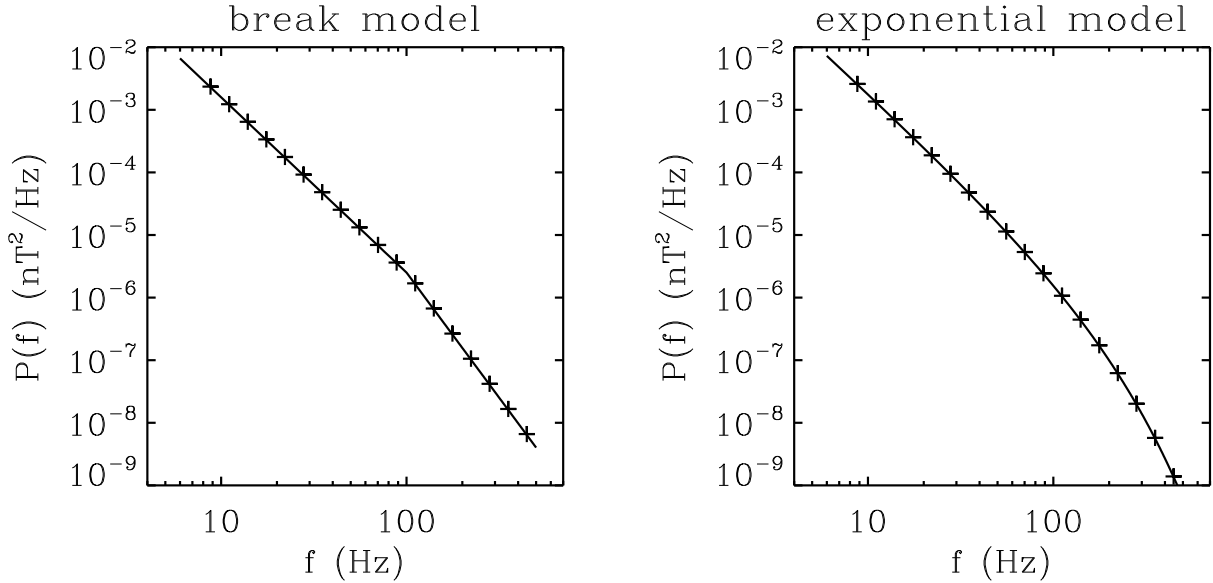}
\caption{\label{fig:a2}  Comparison between an analytic power spectrum $P_A(f)$ (solid line)
and the power spectrum $P_T(f)$ (crosses) calculated by a trapeze integration 
of $P_A(f)$ with the logarithmic frequency binning of the Spectrum Analyser  ($2df=26\%$), for the two models discussed in the paper.
The difference between $P_T(f)$ and $P_A(f)$ is less than 3\%.   } 
\end{figure}

Another argument against the observed spectral bending could be that it is  an artefact due to the logarithmic frequency binning of the Spectrum Analyser  (SA) on Cluster. 

Indeed, the centre frequency of the output channels of SA are 
distributed logarithmically (between 8.8 Hz and 3.56 kHz), and each channel has a bandpass proportional to  its centre frequency, $2 df = 26 f/100$. As the onboard waveforms used by SA are lost, we have no way to check whether a different frequency binning would give different frequency spectra. We only make a crude comparison between an analytical frequency spectrum $P_A(f)$ and the same spectrum integrated by a trapeze method in the logarithmic frequency bands of SA, $P_T(f)$. Note that the trapeze integration has to be made on the logarithms of $P_A(f)$ because the gain of the STAFF-SA receivers is proportional to the logarithm of the power. Figure~\ref{fig:a2} gives the results of this comparison for a spectral break 
model (with two spectral indices $\alpha_1 = 2.8,\; \alpha_2 = 4$, left panel) 
and for an exponential model (right panel) $P_A(f) = f^{-2.7} \exp(-f/f_0)$, with $f_0\simeq 115$~Hz. The solid lines give $P_A(f)$, and the crosses give $P_T(f)$ in the 18 lowest frequencies of SA. It is clear that the trapeze integration in logarithmic channels does not change the shape of the spectra. The ratio $R = P_T/P_A$ is very close to unity: 1.02 to 1.03. There is thus a slight systematic overestimation of the spectrum by the trapeze integration in logarithmic bands. This overestimation (2 to 3\%) cannot increase the downward bending of the spectrum.   We conclude that the logarithmic frequency binning of the Spectrum Analyser cannot smooth a possible spectral break, and cannot produce an artificial downward bending of the spectra.


\bibliographystyle{abbrvnat}
\bibliography{biblio12}

\begin{thebibliography}{43}
\providecommand{\natexlab}[1]{#1}
\providecommand{\url}[1]{\texttt{#1}}
\expandafter\ifx\csname urlstyle\endcsname\relax
  \providecommand{\doi}[1]{doi: #1}\else
  \providecommand{\doi}{doi: \begingroup \urlstyle{rm}\Url}\fi

\bibitem[{Alexandrova}(2008)]{alexandr08npg}
O.~{Alexandrova}.
\newblock {Solar wind vs magnetosheath turbulence and Alfv{\'e}n vortices}.
\newblock \emph{Nonlinear Processes in Geophysics}, 15:\penalty0 95--108, Feb.
  2008.

\bibitem[{Alexandrova} et~al.(2007){Alexandrova}, {Carbone}, {Veltri}, and
  {Sorriso-Valvo}]{alexandr07}
O.~{Alexandrova}, V.~{Carbone}, P.~{Veltri}, and L.~{Sorriso-Valvo}.
\newblock {Solar wind Cluster observations: Turbulent spectrum and role of Hall
  effect}.
\newblock \emph{\planss}, 55:\penalty0 2224--2227, Dec. 2007.
\newblock \doi{10.1016/j.pss.2007.05.022}.

\bibitem[{Alexandrova} et~al.(2008){Alexandrova}, {Carbone}, {Veltri}, and
  {Sorriso-Valvo}]{alexandr08apj}
O.~{Alexandrova}, V.~{Carbone}, P.~{Veltri}, and L.~{Sorriso-Valvo}.
\newblock {Small-Scale Energy Cascade of the Solar Wind Turbulence}.
\newblock \emph{Astrophys. J.}, 674:\penalty0 1153--1157, Feb. 2008.
\newblock \doi{10.1086/524056}.

\bibitem[{Alexandrova} et~al.(2009){Alexandrova}, {Saur}, {Lacombe},
  {Mangeney}, {Mitchell}, {Schwartz}, and {Robert}]{alexandr09}
O.~{Alexandrova}, J.~{Saur}, C.~{Lacombe}, A.~{Mangeney}, J.~{Mitchell}, S.~J.
  {Schwartz}, and P.~{Robert}.
\newblock {Universality of Solar-Wind Turbulent Spectrum from MHD to Electron
  Scales}.
\newblock \emph{Physical Review Letters}, 103\penalty0 (16):\penalty0
  165003--+, Oct. 2009.
\newblock \doi{10.1103/PhysRevLett.103.165003}.

\bibitem[{Alexandrova} et~al.(2010){Alexandrova}, {Saur}, {Lacombe},
  {Mangeney}, {Schwartz}, {Mitchell}, {Grappin}, and {Robert}]{alexandr10}
O.~{Alexandrova}, J.~{Saur}, C.~{Lacombe}, A.~{Mangeney}, S.~J. {Schwartz},
  J.~{Mitchell}, R.~{Grappin}, and P.~{Robert}.
\newblock {Solar wind turbulent spectrum from MHD to electron scales}.
\newblock \emph{12th Int. Solar Wind Conference, AIP Conference Proceedings},
  1216:\penalty0 144--147, Mar. 2010.
\newblock \doi{10.1063/1.3395821}.

\bibitem[{Bale} et~al.(2005){Bale}, {Kellogg}, {Mozer}, {Horbury}, and
  {Reme}]{bale05}
S.~D. {Bale}, P.~J. {Kellogg}, F.~S. {Mozer}, T.~S. {Horbury}, and H.~{Reme}.
\newblock {Measurement of the Electric Fluctuation Spectrum of
  Magnetohydrodynamic Turbulence}.
\newblock \emph{Physical Review Letters}, 94\penalty0 (21):\penalty0 215002--+,
  June 2005.
\newblock \doi{10.1103/PhysRevLett.94.215002}.

\bibitem[{Bale} et~al.(2009){Bale}, {Kasper}, {Howes}, {Quataert}, {Salem}, and
  {Sundkvist}]{bale09}
S.~D. {Bale}, J.~C. {Kasper}, G.~G. {Howes}, E.~{Quataert}, C.~{Salem}, and
  D.~{Sundkvist}.
\newblock {Magnetic Fluctuation Power Near Proton Temperature Anisotropy
  Instability Thresholds in the Solar Wind}.
\newblock \emph{Physical Review Letters}, 103\penalty0 (21):\penalty0
  211101--+, Nov. 2009.
\newblock \doi{10.1103/PhysRevLett.103.211101}.

\bibitem[{Biskamp} et~al.(1996){Biskamp}, {Schwarz}, and {Drake}]{Biskamp96}
D.~{Biskamp}, E.~{Schwarz}, and J.~F. {Drake}.
\newblock {Two-Dimensional Electron Magnetohydrodynamic Turbulence}.
\newblock \emph{Physical Review Letters}, 76:\penalty0 1264--1267, Feb. 1996.
\newblock \doi{10.1103/PhysRevLett.76.1264}.

\bibitem[{Boldyrev} and {Perez}(2012)]{boldyrev12}
S.~{Boldyrev} and J.~C. {Perez}.
\newblock {Spectrum of Kinetic-Alfv{\'e}n Turbulence}.
\newblock \emph{\apjl}, 758:\penalty0 L44, Oct. 2012.
\newblock \doi{10.1088/2041-8205/758/2/L44}.

\bibitem[{Bourouaine} et~al.(2012){Bourouaine}, {Alexandrova}, {Marsch}, and
  {Maksimovic}]{bourouaine12}
S.~{Bourouaine}, O.~{Alexandrova}, E.~{Marsch}, and M.~{Maksimovic}.
\newblock {On Spectral Breaks in the Power Spectra of Magnetic Fluctuations in
  Fast Solar Wind between 0.3 and 0.9 AU}.
\newblock \emph{\apj}, 749:\penalty0 102, Mar. 2012.
\newblock \doi{10.1088/0004-637X/749/2/102}.

\bibitem[{Chen} et~al.(2010){Chen}, {Horbury}, {Schekochihin}, {Wicks},
  {Alexandrova}, and {Mitchell}]{chen10}
C.~H.~K. {Chen}, T.~S. {Horbury}, A.~A. {Schekochihin}, R.~T. {Wicks},
  O.~{Alexandrova}, and J.~{Mitchell}.
\newblock {Anisotropy of Solar Wind Turbulence between Ion and Electron
  Scales}.
\newblock \emph{Physical Review Letters}, 104\penalty0 (25):\penalty0
  255002--+, June 2010.
\newblock \doi{10.1103/PhysRevLett.104.255002}.

\bibitem[{Chen} et~al.(1993){Chen}, {Doolen}, {Herring}, {Kraichnan}, {Orszag},
  and {She}]{chen93}
S.~{Chen}, G.~{Doolen}, J.~R. {Herring}, R.~H. {Kraichnan}, S.~A. {Orszag}, and
  Z.~S. {She}.
\newblock {Far-dissipation range of turbulence}.
\newblock \emph{Physical Review Letters}, 70:\penalty0 3051--3054, May 1993.
\newblock \doi{10.1103/PhysRevLett.70.3051}.

\bibitem[{Cornilleau-Wehrlin et al.}(1997)]{cornilleau97}
N.~{Cornilleau-Wehrlin et al.}
\newblock {The Cluster Spatio-Temporal Analysis of Field Fluctuations (STAFF)
  Experiment}.
\newblock \emph{Space Science Reviews}, 79:\penalty0 107--136, Jan. 1997.

\bibitem[{Escoubet} et~al.(1997){Escoubet}, {Schmidt}, and
  {Goldstein}]{Escoubet1997}
C.~P. {Escoubet}, R.~{Schmidt}, and M.~L. {Goldstein}.
\newblock {Cluster - Science and Mission Overview}.
\newblock \emph{Space Science Reviews}, 79:\penalty0 11--32, Jan. 1997.
\newblock \doi{10.1023/A:1004923124586}.

\bibitem[{Etcheto} and {Faucheux}(1984)]{etcheto84}
J.~{Etcheto} and M.~{Faucheux}.
\newblock {Detailed study of electron plasma waves upstream of the earth's bow
  shock}.
\newblock \emph{J. Geophys. Res.}, 89:\penalty0 6631--6653, Aug. 1984.
\newblock \doi{10.1029/JA089iA08p06631}.

\bibitem[{Filbert} and {Kellogg}(1979)]{Filbert1979}
P.~C. {Filbert} and P.~J. {Kellogg}.
\newblock {Electrostatic noise at the plasma frequency beyond the earth's bow
  shock}.
\newblock \emph{J. Geophys. Res.}, 84:\penalty0 1369--1381, Apr. 1979.
\newblock \doi{10.1029/JA084iA04p01369}.

\bibitem[{Frisch}(1995)]{frisch95}
U.~{Frisch}.
\newblock \emph{{Turbulence. The legacy of A.N. Kolmogorov}}.
\newblock Cambridge: Cambridge University Press, 1995.

\bibitem[{Galtier} and {Bhattacharjee}(2003)]{Galtier2003b}
S.~{Galtier} and A.~{Bhattacharjee}.
\newblock {Anisotropic weak whistler wave turbulence in electron
  magnetohydrodynamics}.
\newblock \emph{Physics of Plasmas}, 10:\penalty0 3065, Aug. 2003.

\bibitem[{Galtier} and {Buchlin}(2007)]{GaltierBuchlin2007ApJ}
S.~{Galtier} and E.~{Buchlin}.
\newblock {Multiscale Hall-Magnetohydrodynamic Turbulence in the Solar Wind}.
\newblock \emph{Astrophys. J.}, 656:\penalty0 560--566, Feb. 2007.
\newblock \doi{10.1086/510423}.

\bibitem[{Galtier} et~al.(2005){Galtier}, {Pouquet}, and {Mangeney}]{galtier05}
S.~{Galtier}, A.~{Pouquet}, and A.~{Mangeney}.
\newblock {On spectral scaling laws for incompressible anisotropic
  magnetohydrodynamic turbulence}.
\newblock \emph{Physics of Plasmas}, 12\penalty0 (9):\penalty0 092310, Sept.
  2005.
\newblock \doi{10.1063/1.2052507}.

\bibitem[{Gary} et~al.(2001){Gary}, {Skoug}, {Steinberg}, and {Smith}]{gary01}
S.~P. {Gary}, R.~M. {Skoug}, J.~T. {Steinberg}, and C.~W. {Smith}.
\newblock {Proton temperature anisotropy constraint in the solar wind: ACE
  observations}.
\newblock \emph{\grl}, 28:\penalty0 2759--2762, July 2001.
\newblock \doi{10.1029/2001GL013165}.

\bibitem[{Goldreich} and {Sridhar}(1995)]{Goldreich1995}
P.~{Goldreich} and S.~{Sridhar}.
\newblock {Toward a theory of interstellar turbulence. 2: Strong alfvenic
  turbulence}.
\newblock \emph{The Astrophysical Journal}, 438:\penalty0 763--775, Jan. 1995.
\newblock \doi{10.1086/175121}.

\bibitem[{Grappin} et~al.(1990){Grappin}, {Mangeney}, and {Marsch}]{Grappin90}
R.~{Grappin}, A.~{Mangeney}, and E.~{Marsch}.
\newblock {On the origin of solar wind MHD turbulence - HELIOS data revisited}.
\newblock \emph{J. Geophys. Res.}, 95:\penalty0 8197--8209, June 1990.
\newblock \doi{10.1029/JA095iA06p08197}.

\bibitem[{Howes} et~al.(2006){Howes}, {Cowley}, {Dorland}, {Hammett},
  {Quataert}, and {Schekochihin}]{howes06}
G.~G. {Howes}, S.~C. {Cowley}, W.~{Dorland}, G.~W. {Hammett}, E.~{Quataert},
  and A.~A. {Schekochihin}.
\newblock {Astrophysical Gyrokinetics: Basic Equations and Linear Theory}.
\newblock \emph{\apj}, 651:\penalty0 590--614, Nov. 2006.
\newblock \doi{10.1086/506172}.

\bibitem[{Howes} et~al.(2011){Howes}, {Tenbarge}, and {Dorland}]{howes11pop}
G.~G. {Howes}, J.~M. {Tenbarge}, and W.~{Dorland}.
\newblock {A weakened cascade model for turbulence in astrophysical plasmas}.
\newblock \emph{Physics of Plasmas}, 18\penalty0 (10):\penalty0 102305, Oct.
  2011.
\newblock \doi{10.1063/1.3646400}.

\bibitem[{Kolmogorov}(1941)]{Kolmogorov1941}
A.~N. {Kolmogorov}.
\newblock {The local structure of turbulence in incompressible viscous fluid
  for very large Reynolds numbers}.
\newblock \emph{Dokl. Akad. Nauk SSSR}, 30:\penalty0 9--13, 1941.

\bibitem[{Lacombe} et~al.(1985){Lacombe}, {Mangeney}, {Harvey}, and
  {Scudder}]{lacombe85}
C.~{Lacombe}, A.~{Mangeney}, C.~C. {Harvey}, and J.~{Scudder}.
\newblock {Electron Plasma Waves Upstream of the Earth's Bow Shock}.
\newblock \emph{J. Geophys. Res.}, 90:\penalty0 73--94, 1985.
\newblock \doi{10.1029/JA090iA01p00073}.

\bibitem[{Leamon} et~al.(1998){Leamon}, {Smith}, {Ness}, {Matthaeus}, and
  {Wong}]{Leamon1998}
R.~J. {Leamon}, C.~W. {Smith}, N.~F. {Ness}, W.~H. {Matthaeus}, and H.~K.
  {Wong}.
\newblock {Observational constraints on the dynamics of the interplanetary
  magnetic field dissipation range}.
\newblock \emph{J. Geophys. Res.}, 103\penalty0 (12):\penalty0 4775, mar 1998.

\bibitem[{Leamon} et~al.(1999){Leamon}, {Smith}, {Ness}, and
  {Wong}]{Leamon1999}
R.~J. {Leamon}, C.~W. {Smith}, N.~F. {Ness}, and H.~K. {Wong}.
\newblock {Dissipation range dynamics: Kinetic Alfv\'en waves and the
  importance of electron beta $\beta_e$}.
\newblock \emph{J. Geophys. Res.}, 104\penalty0 (13):\penalty0 22331--22344,
  Oct. 1999.
\newblock \doi{10.1029/1999JA900158}.

\bibitem[{Leamon} et~al.(2000){Leamon}, {Matthaeus}, {Smith}, {Zank}, {Mullan},
  and {Oughton}]{Leamon2000}
R.~J. {Leamon}, W.~H. {Matthaeus}, C.~W. {Smith}, G.~P. {Zank}, D.~J. {Mullan},
  and S.~{Oughton}.
\newblock {MHD-driven Kinetic Dissipation in the Solar Wind and Corona}.
\newblock \emph{The Astrophysical Journal}, 537:\penalty0 1054--1062, July
  2000.
\newblock \doi{10.1086/309059}.

\bibitem[{Li} et~al.(2001){Li}, {Gary}, and {Stawicki}]{Li2001}
H.~{Li}, S.~P. {Gary}, and O.~{Stawicki}.
\newblock {On the dissipation of magnetic fluctuations in the solar wind}.
\newblock \emph{Geophysical Research Letters}, 28:\penalty0 1347--1350, Apr.
  2001.
\newblock \doi{10.1029/2000GL012501}.

\bibitem[{Matteini} et~al.(2007){Matteini}, {Landi}, {Hellinger}, {Pantellini},
  {Maksimovic}, {Velli}, {Goldstein}, and {Marsch}]{matteini07}
L.~{Matteini}, S.~{Landi}, P.~{Hellinger}, F.~{Pantellini}, M.~{Maksimovic},
  M.~{Velli}, B.~E. {Goldstein}, and E.~{Marsch}.
\newblock {Evolution of the solar wind proton temperature anisotropy from 0.3
  to 2.5 AU}.
\newblock \emph{\grl}, 34:\penalty0 L20105, Oct. 2007.
\newblock \doi{10.1029/2007GL030920}.

\bibitem[{Matteini} et~al.(2011){Matteini}, {Hellinger}, {Landi},
  {Tr{\'a}vn{\'{\i}}{\v c}ek}, and {Velli}]{matteini11}
L.~{Matteini}, P.~{Hellinger}, S.~{Landi}, P.~M. {Tr{\'a}vn{\'{\i}}{\v c}ek},
  and M.~{Velli}.
\newblock {Ion Kinetics in the Solar Wind: Coupling Global Expansion to Local
  Microphysics}.
\newblock \emph{Space Sci. Rev.}, page 128, Apr. 2011.
\newblock \doi{10.1007/s11214-011-9774-z}.

\bibitem[{Narita} and {Gary}(2010)]{narita-gary10}
Y.~{Narita} and S.~P. {Gary}.
\newblock {Inertial-range spectrum of whistler turbulence}.
\newblock \emph{Annales Geophysicae}, 28:\penalty0 597--601, Feb. 2010.
\newblock \doi{10.5194/angeo-28-597-2010}.

\bibitem[{Podesta} and {Gary}(2011)]{podesta-gary11}
J.~J. {Podesta} and S.~P. {Gary}.
\newblock {Magnetic Helicity Spectrum of Solar Wind Fluctuations as a Function
  of the Angle with Respect to the Local Mean Magnetic Field}.
\newblock \emph{\apj}, 734:\penalty0 15, June 2011.
\newblock \doi{10.1088/0004-637X/734/1/15}.

\bibitem[{Sahraoui} et~al.(2010){Sahraoui}, {Goldstein}, {Belmont}, {Canu}, and
  {Rezeau}]{sahraoui10}
F.~{Sahraoui}, M.~L. {Goldstein}, G.~{Belmont}, P.~{Canu}, and L.~{Rezeau}.
\newblock {Three Dimensional Anisotropic k Spectra of Turbulence at Subproton
  Scales in the Solar Wind}.
\newblock \emph{Physical Review Letters}, 105\penalty0 (13):\penalty0
  131101--+, Sept. 2010.
\newblock \doi{10.1103/PhysRevLett.105.131101}.

\bibitem[{Saito} et~al.(2010){Saito}, {Gary}, and {Narita}]{saito10}
S.~{Saito}, S.~P. {Gary}, and Y.~{Narita}.
\newblock {Wavenumber spectrum of whistler turbulence: Particle-in-cell
  simulation}.
\newblock \emph{Physics of Plasmas}, 17\penalty0 (12):\penalty0 122316, Dec.
  2010.
\newblock \doi{10.1063/1.3526602}.

\bibitem[{Salem} et~al.(2012){Salem}, {Howes}, {Sundkvist}, {Bale}, {Chaston},
  {Chen}, and {Mozer}]{salem12}
C.~S. {Salem}, G.~G. {Howes}, D.~{Sundkvist}, S.~D. {Bale}, C.~C. {Chaston},
  C.~H.~K. {Chen}, and F.~S. {Mozer}.
\newblock {Identification of Kinetic Alfv{\'e}n Wave Turbulence in the Solar
  Wind}.
\newblock \emph{\apjl}, 745:\penalty0 L9, Jan. 2012.
\newblock \doi{10.1088/2041-8205/745/1/L9}.

\bibitem[{Schekochihin} et~al.(2009){Schekochihin}, {Cowley}, {Dorland},
  {Hammett}, {Howes}, {Quataert}, and {Tatsuno}]{Schekochihin2009ApJS}
A.~A. {Schekochihin}, S.~C. {Cowley}, W.~{Dorland}, G.~W. {Hammett}, G.~G.
  {Howes}, E.~{Quataert}, and T.~{Tatsuno}.
\newblock {Astrophysical Gyrokinetics: Kinetic and Fluid Turbulent Cascades in
  Magnetized Weakly Collisional Plasmas}.
\newblock \emph{Astrophys. J. Suppl. Ser.}, 182:\penalty0 310--377, May 2009.
\newblock \doi{10.1088/0067-0049/182/1/310}.

\bibitem[{Smith} et~al.(2006){Smith}, {Hamilton}, {Vasquez}, and
  {Leamon}]{smith06}
C.~W. {Smith}, K.~{Hamilton}, B.~J. {Vasquez}, and R.~J. {Leamon}.
\newblock {Dependence of the Dissipation Range Spectrum of Interplanetary
  Magnetic Fluctuationson the Rate of Energy Cascade}.
\newblock \emph{Astrophys. J. Letters}, 645:\penalty0 L85--L88, July 2006.
\newblock \doi{10.1086/506151}.

\bibitem[{Smith} et~al.(2012){Smith}, {Vasquez}, and {Hollweg}]{smith12}
C.~W. {Smith}, B.~J. {Vasquez}, and J.~V. {Hollweg}.
\newblock {Observational Constraints on the Role of Cyclotron Damping and
  Kinetic Alfv{\'e}n Waves in the Solar Wind}.
\newblock \emph{\apj}, 745:\penalty0 8, Jan. 2012.
\newblock \doi{10.1088/0004-637X/745/1/8}.

\bibitem[{Stawicki} et~al.(2001){Stawicki}, {Gary}, and {Li}]{Stawicki2001}
O.~{Stawicki}, S.~P. {Gary}, and H.~{Li}.
\newblock {Solar wind magnetic fluctuation spectra: Dispersion versus damping}.
\newblock \emph{J. Geophys. Res.}, 106:\penalty0 8273--8282, May 2001.
\newblock \doi{10.1029/2000JA000446}.

\bibitem[{Torrence} and {Compo}(1998)]{Torrence1998}
C.~{Torrence} and G.~P. {Compo}.
\newblock {A Practical Guide to Wavelet Analysis.}
\newblock \emph{Bulletin of the American Meteorological Society}, 79:\penalty0
  61--78, Jan. 1998.

\end{thebibliography}


\acknowledgments
We thank the referee for constructive comments. We are also grateful to G.~Howes for useful indications. We thank the team of  Cluster Active Archive. This work was partly supported by the Centre National d'Etudes Spatiales (CNES) and Program National Soleil--Terre (PNST/INSU).

\end{document}